# Adopting the Cybersecurity Concepts into Curriculum: The Potential Effects on Students Cybersecurity Knowledge

Ahmad Mousa Altamimi, Mohammad Azzeh, and Mahmood Albashayreh

Department of Computer Science, Applied Science Private University, Amman, Jordan

{a_altamimi, m.y.azzeh, m_albashayreh}@asu.edu.jo

*Abstract*— The revolution of using the Internet introduces severe security risks that may lead to data leakage and increase the chance of falling victim to security attacks. To defend against such threats, universities have a distinct role in increasing cybersecurity awareness. However, there is currently a definite lack of university curriculums in promoting cybersecurity knowledge; Information Technology (IT) curriculums are no exception. This study examines the effect of adopting cybersecurity concepts on the IT curriculum and determines the potential effect on students' knowledge in terms of cybersecurity practices, level of awareness, and incident response. To this end, a pilot study was first conducted to measure the current level of cybersecurity awareness. The results revealed that students do not have much knowledge of cybersecurity. Thus, a four-step approach was proposed to infuse the relevant cybersecurity topics in five matched courses based on the latest cybersecurity curricular guidelines (CSEC2017). A sample of 42 students (divided identically into experimental and control groups) was carefully selected in a purposive manner without any prior knowledge of cybersecurity. The students in the experimental group were asked to take the five selected courses over five semesters, one course per semester. In each course, both groups went through a pre-test for the infused topics. Then, the experimental group taught the corresponding infused topics, while the control group was not. At the end of each course, a post-test was administered to both groups, and the t-test was conducted. The results found significant differences between marks of prior and post-tests for 11 out of 14 infused topics. These satisfactory results would encourage universities to infuse cybersecurity concepts into their curriculum. The differences between the experimental and control group were almost positive towards satisfying the hypothesis, which confirms the acquired knowledge for the experimental group. Furthermore, the significant accumulative improvements in every two consecutive courses were also tested. We did not find significant continuous improvements over five semesters. The paper concludes by offering recommendations based on results and analysis to promote the level of awareness.

## 1. INTRODUCTION

The Internet has developed immensely, facilitating doing business and providing individuals and organizations with digital communication. Over the numerous advantages offered by the Internet, it is constantly threatened by many risks that often have serious adverse [1]. New digital threats and cyberattacks are coming from new and unexpected sources. Online phishing, social engineering, and malware are just a few examples of cyberattacks [2]. These attacks negatively affect both individuals and the countries' economies. According to [3], it is estimated that cyberattacks' economic impact will increase by around five trillion dollars per year in the next five years. Cyberattacks are getting more sophisticated in the way they misuse and exploit technological advancement. In September 2017, Equifax, a leading USA credit-reporting agency, revealed that American consumers compromised around 143 million personal and financial information in a cybersecurity breach [4].

In fact, many users are unaware of the concept of cybersecurity and how to protect their information. Users often behave in an insecure manner which makes them easy targets for exploitation [5]. Thus, users' understanding of risks and how to protect themselves from cyber-attacks is therefore fundamental in modern life, and spreading awareness among users will give better chances of positively influencing society as a whole [6]. In this regard, researchers argue that universities have a distinct role in building up and developing cybersecurity awareness. This is supported by the fact that learning is becoming more digital than ever [7] due to the COVID-19 pandemic. According to [8], spreading awareness can be done by adopting new technologies, creating awareness inside the university, proposing specialized security courses, and adding security sciences and knowledge to the existing courses.

However, instead of proposing new security courses, efforts have been devoted to proposing guidelines for adopting cybersecurity concepts in the non-security courses to enhance the appropriateness of practice and get better outcomes [9][10][11]. The most notable guideline is CSEC2017 resulted from several computing organizations' joint force (e.g., ACM, IEEE, IFIP, and SIGSEC) and proposed by the Cybersecurity Community in 2017. The guideline defined the cybersecurity discipline and outlined the concepts that include knowledge areas and crosscutting concepts to provide the basis for knowledge areas in cybersecurity. Students will be empowered with the necessary knowledge to act reasonably in various circumstances and deal with their social reality issues [12].

As the first step of our study, we conducted a pilot study on 40 students attending the Information Security course to assess the current level of cybersecurity awareness. The pilot study's main objective was to see how much the students are aware of cyber-attacks and what they do to protect themselves. An online survey is designed specifically to collect data and explore the current cybersecurity behaviour in the following aspects: the general knowledge about cybersecurity and cyber-attacks, password, authentication, email security, firewalls, and mobile security. The results of this survey indicated that students do not have much knowledge of cybersecurity; this lack of knowledge is reflecting when using the Internet while not protecting their data, even on university systems. For example, results reported low levels of authentication usage or password complexity for accounts and how to protect themselves from potential cyber-attacks such as ransomware.

These findings encourage us to carry out our study. Thus, we proposed a four-step methodology to leverage the cybersecurity concepts into non-security computer science courses and assess the potential effects on student's cybersecurity knowledge. Firstly, five principles have been selected from the Cybersecurity Community guideline (CSEC2017) that worked best for the program. The other principles are purposefully not included as they are less important and can be integrated with other concepts. Secondly, the existing curriculum content is mapped to the selected principles. Thirdly, the gaps in the curriculum are identified, where they are filled by infusing new cybersecurity topics. Finally, the gained cybersecurity knowledge is measured to determine the effect of this infusion. The selected security principles are listed in Table 1 along with the related courses, while the complete list of the integrated topics is mentioned in Table 2. It is worth noting that the teaching approach is assumed to not influencing the academic performance of students. The assumption was based on the fact that the universities provide a good teaching approach.

Table 1. The selected concepts and corresponding courses

| Principles | Course | | | | |
|---|---|---|---|---|---|
| | SE | WP | DCN | DB | MP |
| Fault Tolerance | X | | | | |
| Cryptography Algorithms | | X | | | |
| Secure Networking Protocols | | | X | | |
| Authentication Techniques | | | | X | X |
| Hash Functions | | | | X | |

SE = Software Engineering, WP = Web-based Programming, DCN = Data Communication and Networks, DS = Database Systems, MP = Mobile Programming, X = selected.

Table 2. The selected courses and their topics

| Order | Course name | Level | Topics |
|---|---|---|---|
| 1 | Software Engineering (SE) | Year 2/ Semester 1 | SE1: Security breaches |
| | | | SE2: Software vulnerabilities |
| | | | SE3: Fault tolerance techniques |
| 2 | Web-Based Programming (WP) | Year 2/ Semester 2 | WP1: Cryptography |
| | | | WP2: Email and Web Security Protocols |
| | | | WP3: Secure Sockets Layer (SSL) Protocol |
| 3 | Data Communication and Networks (DCN) | Year 3/ Semester 1 | DCN1: Protecting Computing Devices |
| | | | DCN2: Firewalls Types |
| | | | DCN3: Two Factor and Mutual Authentication Techniques |
| 4 | Database Systems (DB) | Year 3/ Semester 2 | DB1: Creating and Managing Passwords |
| | | | DB2: SQL injection Attack |
| | | | DB3: Hash Functions |
| 5 | Mobile Programming (MP) | Year 4/ Semester 1 | MP1: Mobile Breaches |
| | | | MP2: Implementing Security Defenses |

Accordingly, the following hypotheses are proposed:

$H_a$: Infusing cybersecurity principles into non-security courses will improve students' awareness and knowledge of cybersecurity.

$H_b$: Student's cybersecurity awareness and knowledge improvement are directly associated with the course level for the experimental group.

To conduct the experiment, a sample of 42 IT students have been selected with no prior knowledge in cybersecurity. The sample is then divided into two identical groups, 21 students in the experimental group (E) and 21 students in the other control group (C). The students in the experimental group agreed to take the five selected courses in consecutive order, as shown in Figure 1. In each course, all students in both groups were asked to undergo pre-evaluation evaluation tests before enrolling on the selected course, whereas the experimental group is also administrated to another post-evaluation after the end of that course. The questions of both tests have set by expert instructors in the field of cybersecurity, and both tests have a different sample of questions.

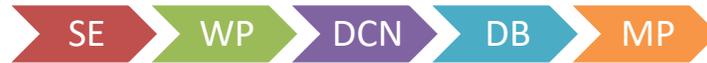

Figure 1. Timeline of the selected course

To statistically test the first hypotheses $H_a$, we performed two kinds of comparisons: 1) within experimental group and 2) across different groups. Concerning within experimental group, the first comparison was conducted to examine the significant difference between the pre-test and post-test scores of the experimental group alone, and control group alone using paired t-test. The second comparison was conducted to examine the significant difference between pre-test and post-test of both experimental and control group, using the two-sample t-test. The purpose of both tests is to make sure that the students in the experimental group are really acquired the required cybersecurity knowledge. On the other hand, to statistically test the second hypothesis $H_b$, we first compute the average marks of the experimental group students integrated topics in each course. Then we conducted a paired t-test statistical test to examine the accumulative improvement over time by comparing post-evaluation marks of each two consecutive courses such as the following pairs:

(SE and WP), (WP and DCN), (DCN and DB) and (DB and MP). Since courses vary in their complexity and importance, the postmarks are not enough to judge the student knowledge improvements over time. Therefore, we multiplied each postmark of each topic with a penalty weight value. This weight is computed from three important factors: complexity, popularity, and associated risks of that topic. Three expert cybersecurity instructors assess each factor. Then these factors are averaged and converted to a weight value between 0 and 1. Finally, the weight value is multiplied by the student postmark to reflect that influence. Further details about this process are discussed in section 3.3.

The obtained results revealed significant differences between marks of pre-evaluation and post-evaluation tests for most infused topics. Moreover, results show that the postmarks are in general higher than pre marks for the experimental group. The results also demonstrate that infusing important cybersecurity topics within other computer science courses can increase students' awareness and knowledge regarding cybersecurity concepts.

The remainder of this paper is organized as follows. The related work is introduced in Section 2. Section 3 presents the research methodology and the experimental work, along with the evaluation measures. The results are discussed then in Sections 4. Finally, Section 5 presents the conclusion and directions for future research.

## 2. Related Work

According to William stalling in his book [13], security education provides users with the necessary skills to perform their duties. Education allows users to know actions that could compromise security, identify possible attack vectors, and report to appropriate personnel. The idea of this approach has been identified in the early 2000s [14] [15]. In this regard, the work of [15] is especially notable, where authors integrate security concepts into existing computer courses. They emphasized this concept's necessity and provided vital suggestions such as security issues that should be discussed throughout the primary and non-major courses in the computer science curriculum to raise awareness of vulnerabilities, threats, and risks. They also suggested that including the security issues as natural aspects is the most effective way to incorporate security issues into the curriculum [15].

Other researchers have further analyzed this integration and proposed models to provide students with the basic computer security principles without the need for professional instructors insecurity [16]. Researchers recently focused on proposing systematic frameworks for proper integration [17][18]. The authors of [18] proposed a framework with three phases structure based on typical curriculum development cycles (e.g., Guideline Development, Planning, and Implementation).

Other studies have also experimented with integrating to determine the students' acquired knowledge [19][20]. For instance, [17] experimented with the integration across low-level courses using security laboratory modules. Results show a positive impact that is reflected in the security knowledge gained by students. In [21], the researchers introduced security teaching with Python and got positive results regarding knowledge and awareness. Unfortunately, the widespread attention to this security integration approach is still insufficient, and its adoption is still minimal [22].

On the other hand, many works have been conducted in the area of educational data mining. Most of them are designed to forecast students' performance to predict the students' results' future outcomes. To this end, historical data were employed (e.g., student GPA, educational background, culture, and academic progress) by many classifications (e.g., Decision Tree, Random Forest, Naïve Bayes, and alike) to get the most accurate prediction [23][24][25]. Here, the GPA was recognized as the most crucial attribute used to predict performance in many works [23].

Student demographics (gender, age, and profession of the parents) is another important attribute used by many works to assess students' final academic performance [26]. Other academic data of students were also used. In addition to grades, significant works considered the online learning platforms as a crucial attribute. The platforms involved MOOCs as EDx and distance learning platforms [27][28][29].

The results are expected to help educators to adopt the proposed security integration approach in their courses. Additionally, the results are expected to encourage researchers to explore further security integration approaches and use other research techniques.

## 3. Research Methodology

The research methodology of our study is divided into three phases. In the first phase, a pilot study is designed based on data collected from 40 students to examine their awareness of cybersecurity concepts. The feedback obtained from the conducted survey enables us to determine security awareness levels that university students already have.

In the second phase, appropriate principles were set and integrated into five non-security courses, following a four-step methodology. Firstly, five appropriate principles have been selected based on the CSEC2017 guideline: fault tolerance, cryptography algorithms, secure networking protocols, authentication techniques, and hash functions. Secondly, a set of courses from the existing curriculum is mapped to the selected principles. Thirdly, the curriculum gaps are identified and filled. Finally, a set of topics is proposed to these gaps.

In the third phase, 42 students have been carefully selected, such that they have no prior knowledge of cybersecurity, nor they have taken any one of the five courses. The students were divided into two identical groups (experimental and control groups). The students in the experimental group agreed to take the five selected courses in consecutive order, as shown in Figure 1. In contrast, the other students in the control group were selected from the registered students in that course. All students were asked to undergo two tests: 1) a pre-evaluation test (pre enrolling on the selected course), 2) a post-test (after the end of that course) on the cybersecurity topics that they have learned within the selected courses. In both tests, the questions were selected carefully by expert instructors in the field of cybersecurity, and both tests have a different sample of questions. A paired t-test statistical test is then performed to examine the significant difference between students' marks in the experimental group for the pre-evaluation and post-evaluation. In addition to using the two-sample t-test to examine the difference between experimental and control groups with respect to pre and post-tests. Finally, we conducted paired t-test to examine the accumulative improvement over time by comparing post-evaluation marks of each two consecutive courses according to the curriculum's timeline. Next, we give a detailed description of these three phases.

### 3.1 The Pilot Study

A pilot study was conducted to get early results and insights into how students are aware of cyber-attacks and how they protect themselves. Hence, an online survey was designed using Microsoft forms 365® to collect quantitative data from 40 students. For privacy reasons, respondents' personally identifying data or any demographic data were not collected in this study except the age range and majors. The respondents are very young, as expected for college students, 18-24 years old. Around 75% are computer science students, while the remaining are software engineering students.

The survey covered different cybersecurity aspects to explore the current state of students' awareness of these aspects: the general knowledge about cybersecurity and cyber-attacks, password, authentication, email security, firewalls, and mobile security. Accordingly, the following questions were formulated:

- On a scale of one to five (five being the most confident), rank your knowledge about cybersecurity and attacks?
- Do you use a strong password to access your social or finical accounts?
- Do you know what Two-Factor Authentication (2FA) is, and do you use it?
- What would you do if you receive an email with links to other sites?
- What would you do when a pop-up window is displayed states that you should download and install a diagnostics program to protect your computer?
- What action do you take if you need to connect to the Internet via an open Wi-Fi hotspot, but it asks you to switch off the firewall?
- Have you ever rejected a mobile app request for accessing your contacts, camera, or location?

## 3.2.   The Selected Principles and Courses

Based on the pilot study results, we found that students have lacked cybersecurity knowledge. Thus, the next step was to assess the potential effects of infusing cybersecurity principles into relevant non-security courses and determine if students' security awareness and knowledge are improved or not. To that end, we relayed on the guideline CSEC2017 for infusing cybersecurity principles in the curriculum. The guideline defined the cybersecurity discipline and outlined the concepts that include knowledge areas and crosscutting concepts to provide the basis for knowledge areas. According to CSEC2017, the curriculum must include the following knowledge areas and concepts: Cryptography algorithms for securing data at rest; Networking security protocols when data is in transit; Availability of Data to be accessible; Authentication and Hash functions to ensure confidentiality and integrity; Security aspects of designing system's component; and Fault tolerance to ensure availability.

Thus, we distilled and selected the following security principles that worked best for the program from these areas and concepts (Fault Tolerance, Cryptography Algorithms, Secure Networking Protocols, Authentication Techniques, and Hash Functions). The others are purposefully not included as they are less important and can be integrated with other concepts. However, besides these principles, eight more concepts have been added to the courses to give students a wide range of cybersecurity topics, from general and fundamental topics to more specific ones [30]. This includes Security breaches, Software vulnerabilities, Protecting Computing Devices, Creating and Managing Passwords, Firewall Types, SQL injection Attack, Mobile Breaches, and Implementing Security Defenses to prevent these breaches.

Once the principles are set, we mapped them to the relevant curriculum courses and proposed a set of topics that will reflect the selected principles. It is important to mention here that participated students were chosen such that they did not take any cybersecurity course before. Below, we illustrate why these specific courses were chosen, including their mapping to principles.

- Software Engineering course (a year 2/semester 1 compulsory course): three principles have been mapped to the Software Engineering course to address the need for system security during the software development life cycle. This is compatible with the course outcomes, where students describe the software engineering processes, activities and distinguish between functional and non-functional requirements. So, students discover the importance of secure system design and how a security-first mindset can improve the software development process by discussing security breaches, software vulnerabilities, fault tolerance techniques, and implicating the software development process.

- Web-Based Programming course (a year 2/ semester 2 compulsory course): after getting the basic cybersecurity principles, we consider securing the web-based systems as they are particularly vulnerable to malicious attacks due to the ability to connect remotely. Thus, during the Web-Based Programming course, students will be empowered with many techniques, protocols, and novel concepts (Cryptography, Email and Web Security Protocols, Secure Sockets Layer (SSL) Protocol) to developing secure web-based solutions. This will enrich the course outcomes as students learn how to create secured World Wide Web pages using HTML5, CCS3, and JavaScript. So, upon completing this course, students will have both technical experiences to deliver innovative solutions for end-users and cybersecurity knowledge to reduce the likelihood of solutions being compromised.

- Data Communication and Networks course (a year 3/ semester 1 compulsory course): Students will get new knowledge and strong skills for using the hardware to protect our systems in their third year. According to the course outcomes, students should describe the network's topology and distinguish among types of switching/routing. So, three appropriate principles have been mapped to here, where students will gain knowledge of how to protect computing devices, how to utilize the firewalls to monitor and control incoming and outgoing network traffic, and how to employ the two factor and mutual authentication techniques to verify the others' legality before any data or information is transmitted.

- Database Systems course (a year 3/semester 2 compulsory course): to be at the forefront of the next big step in the fight against cybercrime. The Database Systems course introduces the security challenges and threats associated with database systems. The course also addresses security issues, such as SQL injection Attack and related current technologies to protect the database systems, such as creating and managing passwords and hash functions. This is compatible with the course outcomes, where students should be able to design different databases and use the Structured Query Language (SQL) to manipulate databases.

- Mobile Programming course (a year 4/ semester 1 compulsory course): finally, to help develop students mobile programming abilities. Students will learn many common and essential mobile security breaches during the Mobile Programming course. Alongside, they also how to implement different security defenses to practice their mobile security knowledge and sharpen their skills. This is compatible with the course outcomes, where students learn the required structures of C# and XAML and the code required to connect them, develop mobile applications, and convert the described problem GUI into suitable GUI components (controls) Xamarin.

## 3.3 Participants and Data Collection

A total of 21 different students (10 females, 11 males) participated in this study over five semesters as an experimental group. The participated students were chosen based on the following criteria: 1) they have no prior knowledge in cybersecurity, 2) they have not taken the selected five courses, and 3) they agreed to take the five selected courses in consecutive order. During each semester, the selected courses' instructors have been asked to select another 21 students as a control group. But it is worth noting that the related cybersecurity concepts have been discussed to the experimental group only.

Moreover, at each course, pre-test and post-test exams were conducted for all students (experimental and control groups), to assess student's knowledge in the selected cybersecurity topics. Based on that, the students' marks in the pre-evaluation and the post-evaluation test were collected (out of 100) for each cybersecurity topics in every course. Then two kinds of comparisons were performed within the course and cross courses. Regarding within course, we applied:

1) A paired t-test statistical test to examine the significant difference between the marks in the pre-evaluation and post-evaluation tests of experimental group for each integrated cybersecurity topics. For each kind of comparison, we also record Cohen's d that measures the effect size of a significance test and win-tie-lose for each pre and postmarks comparison as shown in Algorithm 1.

2) Two-sample t-test to examine the difference between experimental and control groups with respect to pre- and post-tests.

Finally, we conducted paired t-test to examine the accumulative improvement over time by comparing post-evaluation marks of each two consecutive courses according to the curriculum's timeline.

---

**Algorithm 1:** win-tie-lose

Requires:
$win \leftarrow 0$
$tie \leftarrow 0$
$lose \leftarrow 0$
repeat
**foreach** $ID \in dataset$ **do**
    **if** $pos\text{-}mark[ID] > pre\text{-}mark[ID]$ **then**
        $win \leftarrow win + 1$
    **else if** $post\text{-}mark[ID] == pre\text{-}mark[ID]$ **then**
        $tie \leftarrow tie + 1$
    **else**
        $lose \leftarrow lose + 1$
    **end if**
**end**

---

Concerning cross courses, we first compute the average of all experimental group marks overall integrated topics in each course. Then we conducted a paired t-test statistical test to examine the accumulative improvement over time by comparing post-evaluation marks of each two consecutive courses such as the following pairs: (SE and WP), (WP and DCN), (DCN and DB) and (DB and MP). However, because courses vary in their complexity and importance, the use of postmarks are not enough to judge the student knowledge improvements over time. Therefore, we multiplied each postmark of each topic with a weight value. This weight reflects three important factors:

A. **Prevalence** of engaged security terms, where a set of cybersecurity terms is sweeping the media. This includes security breaches, software vulnerabilities, and email security.

B. **Complexity** of understanding the engaged security terms as some of the proposed topics are complex and hard to implement, such as cryptography, two factor and mutual authentication techniques, and hash functions.

C. **Real risk** of that engaged security terms, where users may expose to mobile breaches as they are not familiar with protection methods or fault tolerance techniques.

Each factor is measured in a range of 1 to 10 by three cybersecurity expert instructors. Then these factors are averaged and converted to a weight value between 0 and 1, as shown in equation 1. Finally, the weight value is multiplied by the student mark to reflect that influence.

$$w = \frac{\frac{1}{3}\sum_{i=1}^{3} f_i}{10} \tag{1}$$

Where $w$ is the weight calculated for each security topic, $f_i$ is the factor that influences postmark, which takes value between 1 and 10.

## 4   Results and Discussion

### 4.1   Results of Pilot Study

The general knowledge about cybersecurity and cyber-attacks, password, authentication, email security, firewalls, and mobile security are investigated. The results of the study are summarized in Table 3. Regarding the first question about the general knowledge of cybersecurity and security attacks, it was found that only 10% are strongly knowledgeable. The other questions' results have confirmed the responses to this self-evaluation. Considering this result for IT specialists' respondents, this lack of knowledge is likely to be higher in the general population.

Table 3. Results of Pilot study

| Question | Response | | # | % |
|---|---|---|---|---|
| 1. On a scale of one to five (five being the most confident), rank your knowledge about cybersecurity and attacks? | No idea | | 3 | 7.50% |
| | Hear about | | 10 | 25.00% |
| | Some knowledge | | 16 | 40.00% |
| | Good knowledge | | 7 | 17.50% |
| | Strong knowledge | | 4 | 10.00% |
| | | Total | 40 | 100.00% |
| 2. Do you use a strong password to access your social or finical accounts? | Re-use the same password used in other services | | 16 | 40.00% |
| | Create a password that is as easy as possible to remember | | 8 | 20.00% |
| | Create a very complex password and store it in a manager service | | 10 | 25.00% |
| | Create a new password that is similar to another service | | 4 | 10.00% |
| | Create an entirely new strong password | | 2 | 5.00% |
| | | Total | 40 | 100.00% |
| 3. Do you know what Two-Factor Authentication (2FA) is, and do you use it? | Yes | | 7 | 17.50% |
| | No | | 33 | 82.50% |
| | | Total | 40 | 100.00% |
| 4. What would you do if you receive an email with links to other sites? | Do not click the link | | 14 | 35.00% |
| | Click the links because the email server has already scanned the email | | 21 | 52.50% |
| | Hover the mouse on links to verify the destination URL before clicking | | 5 | 12.50% |
| | | Total | 40 | 100.00% |
| 5. What would you do when a pop-up window is displayed states that you should download and install a diagnostics program to protect your computer? | Download, and install the program | | 26 | 65.00% |
| | Inspect the pop-up windows to verify their validity | | 8 | 20.00% |
| | Ignore the message and close the website | | 6 | 15.00% |
| | | Total | 40 | 100.00% |
| 6. What action do you take if you need to connect to the Internet via an open Wi-Fi hotspot, but it asks you to switch off the firewall? | Connect and switch off the firewall | | 28 | 70.00% |
| | Do not connect to it and keep your firewall | | 8 | 20.00% |
| | Connect to it and establishes a VPN to a trusted server | | 4 | 10.00% |
| | | Total | 40 | 100.00% |
| 7. Have you ever rejected a mobile app request for accessing your contacts, camera, or location? | Yes | | 24 | 60.00% |
| | No | | 16 | 40.00% |
| | | Total | 40 | 100.00% |

The second question asked about creating strong passwords. Surprisingly, more than 40% of the respondents tended to re-use the same password for other services, and another 20% preferred to create an easy password. This might be a bad indicator of password creating knowledge. Similarly, the authentication techniques knowledge is not much better, as seen in the results of question 3. Most of the participants do not know what Two-Factor Authentication is.

Regarding the website trust in question 4, the same issue is revealed for respondents who consider the email server responsible for scanning the email links, which is not the case in practice. This awareness level is also reflected in question 5, where around 65% of the participants will download and install a program suggested by another site.

Another parameter that still illustrates low awareness of cybersecurity is shown when 70% of respondents willing to switch off their firewalls for a free Wi-Fi hotspot given in question 6. It is also important to underline that the awareness about denying a mobile app request personal data positively impacts participants' responses. For the last question, 60% of the participants will reject a mobile app request accessing their contacts, camera, or locations.

The results of this survey indicated that students do not have much knowledge of cybersecurity; they need to be motivated to security precautions and be exceptionally the risks of online services. Also, it appears that educational institutions do not have an active approach to improve awareness among students. It is worth mentioning here that our pilot results are compatible with recent studies. One can consider here the study in [31], which analyzed cybersecurity awareness among education sector members in the Middle East region. The results reveal that the participants do not have the requisite knowledge and understanding of the importance of security principles and their practical application in day-to-day work.

## 4.2 Results of Pre-evaluation and Post-evaluation tests

Two types of tests were conducted to statistically analyze the difference between the pre and post-tests marks. While the paired t-test was used to compare the pre and post-test marks for the experimental group, the Two-sample t-test was used to compare between the experimental and control groups in terms of the pre and post-test marks.

### 4.2.1 Statistical Tests Between Pre and Post Exams for Experimental Group.

In these tests, we compared experimental group students' marks pre-starting the course and after finishing the course for each course's infused cybersecurity topic. The average of marks for each cybersecurity topic is converted to a scale from 0 to 100. By integrating additional cybersecurity topics within the software engineering course, we noticed that the students' overall average marks in the post-evaluation test are higher than the pre-evaluation test with significant t-test results for two topics (SE1 & SE2), as shown in Table 4. Surprisingly, we can also notice that pre-evaluation marks' standard deviation is close to that of post-evaluation marks in most courses. The Cohen's d measure indicates the effect size for topics (SE1 & SE2) is greater than 0.5, which means a significant difference between pre and postmarks, which was less than 0.5 for SE1.

Moreover, the number of wins, tie and losses show that the number of wins is greater than losses, indicating that most students obtained better marks after completed infused topics. Figure 2 shows boxplots of pre and postmarks for three infused topics with the software engineering course. We can see that there is no overlap between boxes of pre-evaluation and post-evaluation for the first two cybersecurity topics (SE1 & SE2), while there is a slight overlap for Malware Types and Symptoms (SE3). On the other hand, we found that boxes length of post-evaluation marks in SE1 and SE3 topics are larger than pre-evaluation marks. This indicates that there is large distribution in students' postmarks.

Table 4. Results of Software Engineering Course, using paired t-test

| ID | Cybersecurity Topic | Pre-Test | Post-Test | t-test | Effect size (Cohen's d) | win | tie | lose |
|----|---------------------|----------|-----------|--------|-------------------------|-----|-----|------|
| SE1 | SB | $46.7 \pm 11.4$ | $66.4 \pm 12.8$ | t = -9.0, p-value <0.001* | 1.63 | 19 | 0 | 2 |
| SE2 | SV | $48.3 \pm 18.1$ | $72.7 \pm 13.7$ | t = -8.5, p-value <0.001* | 1.53 | 17 | 0 | 4 |
| SE3 | MTS | $61.0 \pm 15.6$ | $65.7 \pm 13.4$ | t = -1.8, p-value = 0.08 | 0.32 | 12 | 1 | 8 |

* significant at 95%. SB = Security breaches, SV = Software vulnerabilities, MTS = Malware types and symptom.

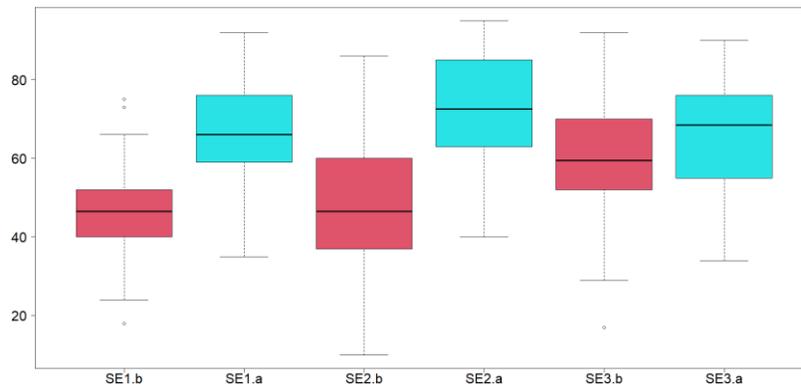

Figure 2. Boxplots of pre and postmarks for SE infused topics (b: before taking the course, a: after completing the course)

The same findings can be seen for the Web Programming course. We adopt three cybersecurity topics, as shown in Table 5. Surprisingly, the Web Programming course's average marks are less than that of the Software Engineering course. The paired t-test between the two evaluations shows significant differences between the two marks for all topics. The Cohen's d effect size confirms the obtained statistical differences with an effect size greater than 0.5. Also, the number of wins is significantly greater than the number of losses, which revealed that the number of students who improved their mark is larger than those who failed to improve his mark. Although the paired t-test showed significant differences between the two tests for Email and Web Security topics (WP2), we can see a large overlap between both evaluation tests, as shown in Figure 3. We can notice that the distribution of postmarks of WP1 is narrower than pre marks, whereas the opposite can be seen for WP2. This confirms that every student's individual capability is different, but it greatly affects the course's overall performance.

Table 5. Results of Web Programming Course, using paired t-test

| ID | Cybersecurity Topic | Before | After | t-test | Effect size (Cohen's d) | win | tie | lose |
|----|---------------------|--------|-------|--------|-------------------------|-----|-----|------|
| WP1 | CRP | $41.1 \pm 14.5$ | $52.5 \pm 9.4$ | t = -5.2394, p-value <0.001* | 0.95 | 14 | 0 | 7 |
| WP2 | EWS | $44.8 \pm 11.9$ | $57.6 \pm 15.9$ | t = -5.0848, p-value <0.001* | 0.92 | 15 | 0 | 6 |
| WP3 | SSL | $30.4 \pm 15.3$ | $46.5 \pm 13.7$ | t = -6.1875, p-value <0.001* | 1.11 | 15 | 1 | 5 |

* significant at 95%. CRP = Cryptography, EWS = Email and web security, SSL = Secure sockets layer (SSL)

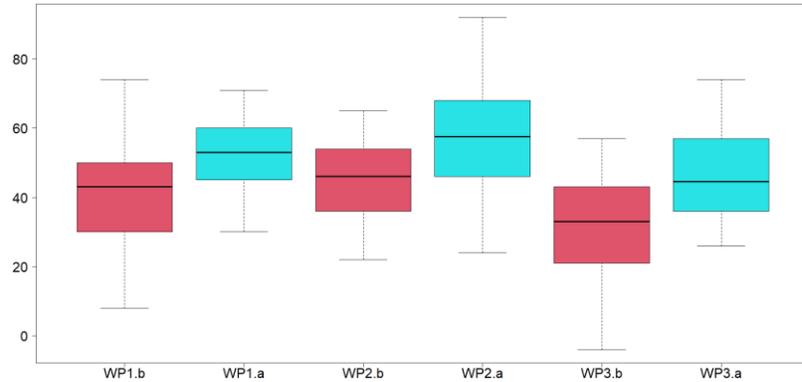

Figure 3. Boxplots of pre and postmarks for WP infused topics (b: before taking the course, a: after completing the course)

The results in Table 6 demonstrate that adopting three cybersecurity topics would partially enhance student awareness regarding the Data communication and networking course. However, only two topics (DCN2 & DCN3) show significant improvements, as confirmed by the t-test. Surprisingly, the average of pre and postmarks for DCN1 are similar with insignificant difference between them. However, the distribution of postmarks is slightly larger than pre marks for DCN1, as shown in Figure 4. In contrast, the overall average of post-evaluation marks is higher than the average of pre-evaluation marks for DCN1 and DCN2. Notably, the post-evaluation's standard deviation is smaller than the pre-evaluation test only for DCN1, suggesting less variability in marks distribution.

However, the obtained Cohen's d effect size reported that the difference between each pair of marks is negligible, even if it is statistically significant. Also, the number of losses is a little high in comparison with the previous course. This shows that the number of students who did not improve their marks is relatively close to those who successfully improved their marks. On the other hand, Figure 4 shows overlaps between each pair of boxes (pre and postmarks). Despite that overlap, the difference between both evaluation marks is significant for two topics. Remarkably, there are no outliers in the three topics, which confirm that all marks fall within the first and third quartile. Thus, in general, we can see satisfactory improvements in student knowledge in DCN course, but we believe that this improvement did not show the expected level that we are looking for.

Table 6. Results of Data Communication and Networking Course, using paired t-test

| ID | Cybersecurity Topic | Before | After | t-test | Effect size (Cohen's d) | win | tie | lose |
|---|---|---|---|---|---|---|---|---|
| DCN1 | PCD | $54.1 \pm 15.1$ | $54.8 \pm 17.5$ | t = -0.24, p-value = 0.814 | 0.04 | 11 | 0 | 10 |
| DCN2 | FT | $50.0 \pm 19.1$ | $57.3 \pm 13.6$ | t = -2.40, p-value = 0.019* | 0.44 | 13 | 0 | 8 |
| DCN3 | TFMA | $42.4 \pm 16.0$ | $48.6 \pm 18.3$ | t = -2.01, p-value = 0.047* | 0.36 | 12 | 1 | 8 |

* significant at 95%. PCD = Protecting computing devices, FT = Firewall types, TFMA = Two factor, and mutual authentication.

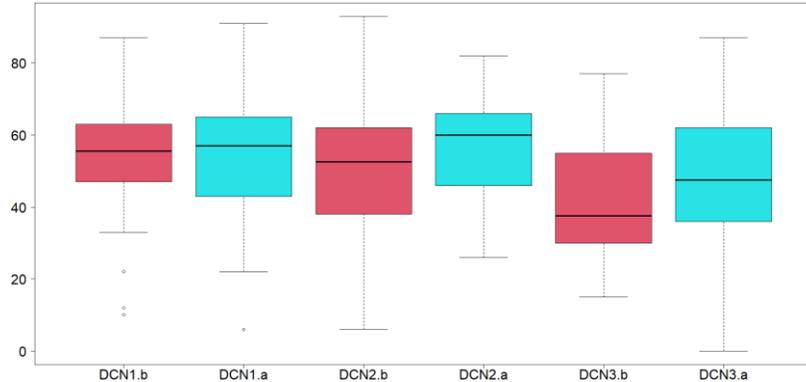

Figure 4. Boxplots of pre and postmarks for DCN infused topics (b: before taking the course, a: after completing the course)

The results of significance tests for the Database course are a little bit different than previous courses. Three cybersecurity topics were adopted, as mentioned in Table 7. The average of marks for post-test is larger in general than the pre-evaluation test, suggesting good improvements in students' awareness. The paired t-test results demonstrate a significant difference between pre-evaluation and post-evaluation marks for three cybersecurity topics: Creating and Managing Passwords and Hash functions. This result suggests that the student awareness significantly improved while taking these topics within the database course. However, the effect size revealed a strong justification to judge that the difference between two evaluation marks is significant for only two topics (DB2 and DB3) with Cohen's d over 0.5.

On the other hand, the number of wins is larger than the number of losses for all topics, confirming significant improvements. Figure 5 also confirms the obtained results, which shows that the difference between every pair of distributions is significant despite a little overlap. However, the boxplots also show a little overlap between both evaluation tests across three topics. In addition, we can see medinas of both tests are quite similar for the first infused cybersecurity topics denoted by DB1. The post-evaluation test's box length is higher than the pre-test for DB1, demonstrating high variability in student marks. This informs us that students' marks in the pre-evaluation test are quite scattered and have a large standard deviation. The opposite can be seen for DB2 and DB3 topics.

Table 5. Results of Database Systems Course, using paired t-test

| ID | Cybersecurity Topic | Before | After | t-test | Effect size (Cohen's d) | win | tie | lose |
|----|---------------------|--------|-------|--------|-------------------------|-----|-----|------|
| DB1 | CMP | $60.7 \pm 11.9$ | $66.3 \pm 12.6$ | t = -2.6, p-value = 0.01* | 0.46 | 13 | 1 | 7 |
| DB2 | SIA | $49.8 \pm 16.4$ | $64.8 \pm 8.80$ | t = -6.4, p-value <0.001* | 1.19 | 16 | 0 | 5 |
| DB3 | HF | $38.6 \pm 15.4$ | $45.4 \pm 10.1$ | t = -2.9, p-value = 0.004* | 0.54 | 14 | 0 | 7 |

* significant at 95%. CMP: Creating and Managing Passwords, SIA: SQL injection Attack, HF: Hash functions

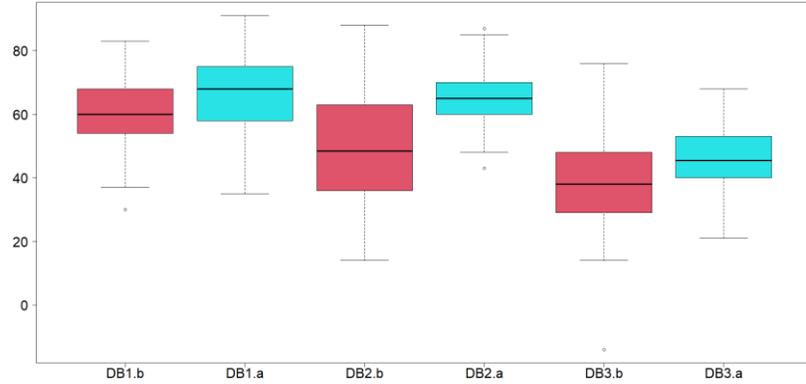

Figure 5. Boxplots of pre and postmarks for DB infused topics (b: before taking the course, a: after completing the course)

Concerning the Mobile Programming course, we can notice from Table 8 that there are significant differences between pre-evaluation and post-evaluation tests for only MP2 topics, confirming that adopting these cybersecurity topics in a programming course would enhance student awareness about threats that can affect the mobile application. In contrast, we did not find this significant difference for MP1. Both findings are confirmed by Cohen's d effect size, which is less than 0.5 for MP1 and greater than 0.5 for MP2. We can also notice that the average marks of pre-evaluation tests for both engaged cybersecurity topics are quite acceptable, demonstrating that the student in this course is familiar with this kind of threat. Also, we did not notice significant improvements in their marks after adopting MP1, which is also confirmed by the number of wins and losses for MP1 that is so close. In contrast, the number of wins for MP2 was significantly larger than losses. These results are also confirmed by boxplots in Figure 6, demonstrating a significant difference between two evaluation marks for only MP2.

Table 8. Results of Mobile Programming Course, using paired t-test

| ID | Cybersecurity Topic | Before | After | t-test | Effect size (Cohen's d) | win | tie | lose |
|---|---|---|---|---|---|---|---|---|
| MP1 | MB | 67.1 ± 11.2 | 70.2 ± 15.4 | t = -1.3, p-value = 0.20 | 0.23 | 12 | 0 | 9 |
| MP2 | ISD | 53.5 ± 12.7 | 73.7 ± 13.4 | t = -8.6, p-value <0.001* | 1.54 | 19 | 0 | 2 |

* significant at 95%. MB = Mobile Breaches, ISD = Implementing Security Defenses.

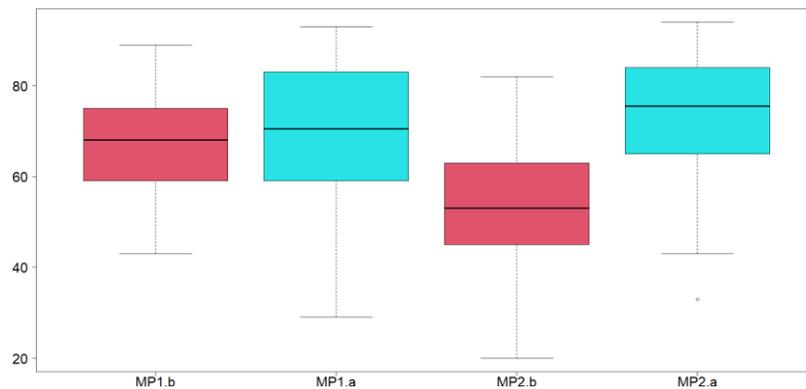

Figure 6. Boxplots of pre and postmarks for MP infused topics (b: before taking the course, a: after completing the course)

### 4.2.2 Statistical test between experimental and control groups

We compared the pre and post-test marks for both control and experimental groups for each course's infused cybersecurity topic in these tests. The average of marks for each cybersecurity topic is converted to a scale from 0 to 100. Table 9 shows the statistical analysis using the Two-sample t-test between the experimental and control group for Software Engineering Course. Results show no significant difference for the pre-test of all software engineering topics between the experimental and control groups. This confirms that students' knowledge in both groups is relatively similar with no significant difference. In contrast, we noticed positive significance for the post-test between the experimental and control group with a large effect size. These findings are consistent with our basic assumptions that presume that the student's marks in the experimental and control group must be relatively similar to the pre-test because they have no prior knowledge and significantly different in terms of post-tests.

Table 9. Comparison between experimental and control group for Software Engineering Course, using Two-sample t-test

| Test Type | ID | Cybersecurity Topic | Experimental Group | Control Group | t-test | Effect size (Cohen's d) |
|---|---|---|---|---|---|---|
| Pre-Test | SE1 | SB | 46.7 ± 11.4 | 51.6 ± 12.3 | t = -1.44, p-value =0.19 | 0.41 |
| | SE2 | SV | 48.3 ± 18.1 | 46.7 ± 11.8 | t = 0.34, p-value =0.74 | 0.10 |
| | SE3 | MTS | 61.0 ± 15.6 | 57.7 ± 14.6 | t = 0.71, p-value = 0.48 | 0.21 |
| Post-Test | SE1 | SB | 66.4 ± 12.8 | 52.4 ± 12.9 | t = 3.53, p-value =0.001* | 1.09 |
| | SE2 | SV | 72.7 ± 13.7 | 48.3 ± 10.2 | t = 6.54, p-value <0.001* | 2.02 |
| | SE3 | MTS | 65.7 ± 13.4 | 55.5 ± 15.2 | t = 2.31, p-value = 0.02* | 0.71 |

\* significant at 95%. SB = Security breaches, SV = Software vulnerabilities, MTS = Malware types and symptom.

Table 10 shows the results for the Web Programming Course. Here, we show a significant difference between both groups regarding the pre-test for the Secure sockets layer (SSL) topic. However, no difference was shown for the remaining topics between the two groups. In terms of post-test, no significant difference was shown between the two groups for the Cryptography topic. This is due to the difficulty of this topic as it depends on complex math theory. However, a significant difference was shown for the remaining topics (e.g., Email and web security and secure sockets layer).

Table 10. Comparison between experimental and control group for Web Programming Course, using Two-sample t-test

| Test Type | ID | Cybersecurity Topic | Experimental Group | Control Group | t-test | Effect size (Cohen's d) |
|---|---|---|---|---|---|---|
| Pre-Test | WP1 | CRP | 41.1 ± 14.5 | 46.3 ± 11.2 | t = -1.30, p-value =0.20 | 0.40 |
| | WP2 | EWS | 44.8 ± 11.9 | 43.6 ± 10.5 | t = 0.347, p-value =0.73 | 0.11 |
| | WP3 | SSL | 30.4 ± 15.3 | 40.7 ± 16.7 | t = -2.08, p-value =0.044* | 0.64 |
| Post-Test | WP1 | CRP | 52.5 ± 9.4 | 48.6 ± 11.7 | t = 1.2, p-value =0.24 | 0.41 |
| | WP2 | EWS | 57.6 ± 15.9 | 47.3 ± 12.3 | t = 2.35, p-value =0.025* | 0.72 |
| | WP3 | SSL | 46.5 ± 13.7 | 38.2 ± 10.1 | t = 2.23, p-value =0.032* | 0.69 |

\* significant at 95%. CRP = Cryptography, EWS = Email and web security, SSL = Secure sockets layer (SSL)

Table 11 shows the results for the Data Communication and Networking Course. For the pre-test marks, no significant difference was shown between both groups for all topics. However, a significant difference was shown in the post-test marks for two topics (e.g., Protecting computing devices and Firewall types). On the other hand, students did not perform well in the Two factor and mutual authentication topic, which is demonstrated by students' marks.

Table 11. Comparison between experimental and control group for Data Communication and Networking Course, using Two-sample t-test

| Test Type | ID | Cybersecurity Topic | Experimental Group | Control Group | t-test | Effect size (Cohen's d) |
|---|---|---|---|---|---|---|
| Pre-Test | DCN1 | PCD | 54.1 ± 15.1 | 52.3 ± 12.4 | t = 0.42, p-value = 0.68 | 0.13 |
| | DCN2 | FT | 50.0 ± 19.1 | 51.7 ± 16.3 | t = -0.31, p-value = 0.76 | 0.10 |
| | DCN3 | TFMA | 42.4 ± 16.0 | 46.1 ± 13.4 | t = -0.81, p-value = 0.42 | 0.25 |
| Post-Test | DCN1 | PCD | 54.8 ± 17.5 | 45.0 ± 12.2 | t = 2.11, p-value = 0.04* | 0.65 |
| | DCN2 | FT | 57.3 ± 13.6 | 46.9 ± 12.1 | t = 2.62, p-value = 0.01* | 0.81 |
| | DCN3 | TFMA | 48.6 ± 18.3 | 41.1 ± 13.4 | t = 1.52, p-value = 0.13 | 0.47 |

\* significant at 95%. PCD = Protecting computing devices, FT = Firewall types, TFMA = Two factor, and mutual authentication.

Table 12 shows the results for the Database Systems Course. The gained results of this course are similar to the previous course (e.g., Data Communication and Networking). Specifically, results show no significant difference between both groups regarding the pre-test for all topics. In addition, in terms of post-test, a significant difference was shown between the two groups for all topics. This is due to the popularity of these topics, as most students used the topics' techniques daily (e.g., Creating and Managing Passwords, SQL injection Attack, and Hash functions).

Table 6. Comparison between experimental and control group for Database Systems Course, using Two-sample t-test

| Test Type | ID | Cybersecurity Topic | Experimental Group | Control Group | t-test | Effect size (Cohen's d) |
|---|---|---|---|---|---|---|
| Pre-Test | DB1 | CMP | 60.7 ± 11.9 | 64.9 ± 13.3 | t = -1.1, p-value = 0.29 | 0.33 |
| | DB2 | SIA | 49.8 ± 16.4 | 51.4 ± 11.8 | t = -0.36, p-value = 0.72 | 0.11 |
| | DB3 | HF | 38.6 ± 15.4 | 42.6 ± 12.1 | t = -0.94, p-value = 0.36 | 0.29 |
| Post-Test | DB1 | CMP | 66.3 ± 12.6 | 57.1 ± 12.4 | t = 2.38, p-value = 0.02* | 0.74 |
| | DB2 | SIA | 64.8 ± 8.80 | 52.3 ± 14.6 | t = 3.36, p-value =0.002* | 1.04 |
| | DB3 | HF | 45.4 ± 10.1 | 38.7 ± 9.5 | t = 2.21, p-value = 0.03* | 0.68 |

\* significant at 95%. CMP: Creating and Managing Passwords, SIA: SQL injection Attack, HF: Hash functions

Table 13 shows the results for the Mobile programming Course. Due to the topic novelty, most of the topics did not significantly differ between the experimental and control groups. For instance, the Mobile Breaches and Implementing Security Defenses topics have no significant difference between groups in pre-test marks. We notice the same case for Implementing Security Defenses post-test marks, where no significant difference was noticed between groups. However, a significant difference was shown for the Mobile Breaches topics, where students gain knowledge after Infusing this principle.

Table 7. Comparison between experimental and control group for Mobile programming Course, using Two-sample t-test

| Test Type | ID | Cybersecurity Topic | Experimental Group | Control Group | t-test | Effect size (Cohen's d) |
|---|---|---|---|---|---|---|
| Pre-Test | MP1 | MB | 67.1 ± 11.2 | 64.3 ± 12.4 | t = 0.77, p-value = 0.44 | 0.24 |
| | MP2 | ISD | 53.5 ± 12.7 | 56.7 ± 11.9 | t = -0.84, p-value =0.40 | 0.26 |
| Post-Test | MP1 | MB | 70.2 ± 15.4 | 61.5 ± 10.3 | t = 2.15, p-value = 0.04* | 0.66 |
| | MP2 | ISD | 73.7 ± 13.4 | 59.9 ± 12.1 | t = 3.5, p-value =0.001 | 1.08 |

\* significant at 95%. MB = Mobile Breaches, ISD = Implementing Security Defenses.

Indeed, from the above statistical test results, we noticed that, in general, the students have a quite low level of cybersecurity awareness, as confirmed in the averages and standard deviations of pre-evaluation marks. But these marks are significantly improved in almost all topics except for three topics, namely, SE3, DCN1, MP1. Therefore, we revisit our first hypothesis:

$H_a$: Infusing cybersecurity principles into non-security courses will improve students' awareness and

knowledge of cybersecurity.

To test this hypothesis, we compute average marks for all pre marks of all topics (Say before) and average post marks for all topics (Say after) for the experimental group only. The paired t-test results (t= -17.3, p-value<0.001) between the before and after groups confirmed Ha's hypothesis and revealed a significant difference between students' average marks in all topics.

Furthermore, we compute average marks for all postmarks of all topics for the experimental group (E) and average postmarks for the control group (C). The we applied two-sample t-test results (t= 2.16, p-value=0.031) between them. The obtained results also confirmed Ha's hypothesis and revealed a significant difference between students' average marks in all topics.

From these results, we encourage higher education institutes to integrate some important cybersecurity topics within existing courses instead of adding new courses or programs. This adoption will eventually reduce security threats in different aspects of our lives, increase students' awareness of various cybersecurity topics, and enrich their knowledge.

## 4.3    Results of Cross Courses Comparison

This section attempts to test the second hypothesis:

$H_b$: Student's cybersecurity awareness and knowledge improvement are directly associated with the level of the course.

If a significant increment in the postmarks is seen along the timeline shown in Figure 1, this hypothesis can be confirmed. To test this hypothesis, we follow the same vein employed with hypothesis 1. Specifically, the average of postmarks of all topics in each course is calculated first. However, since each topic varies in its popularity and complexity, we multiply each topic's postmark with a weight value calculated based on three factors (Prevalence, Complexity and Risk) determined by three cybersecurity expert instructors. Each factor is assessed by a 10-points scale (from 1 to 10) that represents agreed experts' opinion about the strength of that factor in the given topic. The average of three factors' values is then computed for each given topic and converted to a penalty weight, as illustrated in equation 1. The penalty weights are multiplied by students postmarks of each topic to compute weight postmarks. Finally, each course's postmarks are computed by calculating the integrated topics' average weighted postmarks. Table 14 described the given factor values and computed penalty weight for each cybersecurity topic.

Table 8. Factor assessment and weight computation for each cybersecurity topic

| Cybersecurity code | Acronym | Prevalence | Complexity | Real Risk | Average | weight |
|---|---|---|---|---|---|---|
| SE1 | SB | 10 | 7 | 10 | 9.00 | 0.90 |
| SE2 | SV | 9 | 6 | 8 | 7.67 | 0.77 |
| SE3 | MTS | 6 | 6 | 7 | 6.33 | 0.63 |
| WP1 | CRP | 8 | 9 | 10 | 9.00 | 0.90 |
| WP2 | EWS | 8 | 8 | 8 | 8.00 | 0.80 |
| WP3 | SSL | 6 | 9 | 8 | 7.67 | 0.77 |
| DCN | PCD | 8 | 5 | 8 | 7.00 | 0.70 |
| DCN | FT | 3 | 7 | 9 | 6.33 | 0.63 |
| DCN | TFMA | 2 | 8 | 8 | 6.00 | 0.60 |
| DB1 | CMP | 9 | 5 | 9 | 7.67 | 0.77 |
| DB2 | SIA | 8 | 7 | 9 | 8.00 | 0.80 |
| DB3 | HF | 4 | 8 | 8 | 6.67 | 0.67 |
| MP1 | MB | 7 | 9 | 9 | 8.33 | 0.83 |
| MP2 | ISD | 5 | 8 | 8 | 7.00 | 0.70 |

According to the curriculum, the five selected topics should be taken in a restricted order, as illustrated in Table 2 and Figure 1, distributed over five semesters. The participated students can take one course each semester. To evaluate the accumulative improvements, we proposed to measure the significance t-test between every two consecutive courses based on the weighted postmarks. According to Table 2, There are four possible consecutive courses that we should test: (SE and WP), (WP and DCN), (DCN and DB) and (DB and MP). The knowledge improvement means that the weighted student postmarks are supposed to increase significantly over time. Table 15 shows the t-test for each consecutive courses. We can see that all pair of consecutive courses' differences are significant, but the weighted postmarks did not improve over time, as shown in the boxplots in Figure 7. The courses are sorted in the predefined order from left to right. For example, we can see that the average of weighted postmarks for WP is less than that of SE. The same pattern can be found for the next consecutive courses (WP & DCN). However, we can see improvements in weighted postmarks after DCN course. Finally, we could not confirm the second hypothesis because we could not find a significant increment in students weighted postmarks for each consecutive courses.

Table 9. t-test between each pair of consecutive courses, using paired t-test

| Pair of courses | Precedent mean | Antecedent mean | Status | t-test | Cohen's d effect size | win | tie | loss |
|---|---|---|---|---|---|---|---|---|
| SE vs. WP | 52.4 ± 5.9 | 43.1 ± 6.3 | deterioration | t = 8.51, df = 61, p-value < 0.001 | 1.53 | 11 | 0 | 51 |
| WP vs. DCN | 43.1 ± 6.3 | 34.5 ± 6.0 | deterioration | t = 7.69, df = 61, p-value < 0.001 | 1.40 | 10 | 0 | 52 |
| DCN vs. DB | 34.5 ± 6.0 | 44.5 ± 4.3 | Improvement | t = 10.5, df = 61, p-value < 0.001 | 1.94 | 54 | 0 | 8 |
| DB vs. MP | 44.5 ±4.3 | 54.9 ± 7.7 | Improvement | t = 16.5, df = 61, p-value < 0.001 | 1.73 | 53 | 0 | 9 |

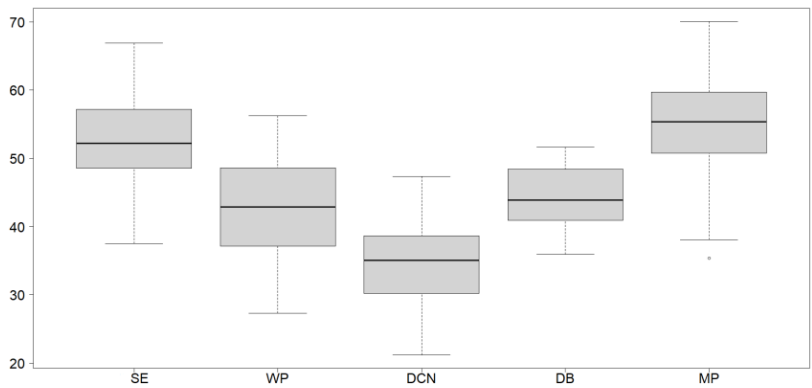

Figure 7. Boxplots of weighted post marks for five selected courses

## 5 THREATS TO VALIDITY

The threats to this study's validity can be seen from three dimensions: 1) the population, 2) the selected courses, and 3) the proposed cybersecurity topics. Regarding the selected population, we have selected 62 students from different years and two IT disciplines. The choice of IT students only and neglecting other university students might threaten the validity of this study. However, we chose IT students for two reasons: first, the study aims to leverage non-cyber security courses in computer science curricula with cybersecurity terms, which fits mainly with IT courses. Secondly, some of the proposed terms need experience in IT. The sample size was 62, and we believe it is sufficient to conduct the experiment and perform statistical analysis. Another threat that might threaten the validity is choosing only one university. We believe that the selected sample represents the population of IT students in most universities. Also, we tried to make the number of females and males relatively equal, but unfortunately, it was hard because the majority of students in the selected courses are male. Concerning the selected courses, we have selected five courses that match best with the infused cybersecurity principles. We believe that the cybersecurity topics are quite related to the content of these courses.

On the other hand, to decrease the bias of factors assessment that is used to compute penalty, we asked three expert cybersecurity instructors, after an agreement, to assess each topic. Also, the questions of pre-test and post must show different complexity levels. Therefore, all questions have been written carefully such that the questions of post-evaluation are a little complex than the prior evaluation test. Finally, the cybersecurity topics have been selected from a well-known guideline that is proposed in CSEC2017. Therefore, these topics are well recognized and reflect the industrial needs of IT students.

## 6  CONCLUSION

This paper proposes a detailed approach for examining the effect of infusing cybersecurity principles in the IT curriculum's non-security courses on students' awareness and cybersecurity knowledge. However, before determining that, we started our study with a pilot study conducted on 40 IT students to see how much students are aware of 7 principles related to cybersecurity and what they do to protect themselves from cyber-attacks. The obtained results indicated that students do not have much knowledge of cybersecurity and need to be aware of security precautions and online services risks. Also, results revealed that educational institutions do not actively approach cybersecurity awareness among students. Based on this finding, we relied on the remarkable guideline (CSEC2017) and distilled the main security principles that the curriculum must include. Accordingly, these principles are mapped to the relevant curriculum courses and proposed a set of topics that will reflect the selected principles. To determine the effects of infusing principles, we assessed the degree of improvements in the acquired knowledge for 42 students through pre and post-evaluation tests. The students were divided into two identical groups (experimental and control groups). All students were asked to undergo two tests, a pre-evaluation test (pre enrolling on the selected course) and a post-test (after the end of that course) on the cybersecurity topics. In both tests, the questions were selected carefully by expert instructors in cybersecurity, and both tests have a different sample of questions. A paired t-test statistical test is then performed to examine the significant difference between experimental group students' marks in the pre-evaluation and post-evaluation. In addition, the two-sample t-test is used to examine the difference between experimental and control groups for pre and post-tests. We noticed that the students often have a quite low level of cybersecurity awareness, as confirmed in the averages and standard deviations of pre-evaluation marks. However, the pre-evaluation marks are significantly improved in 11 out of 14 topics. Moreover, results show that the postmarks are in general higher than pre marks. The results also demonstrate that engaging important cybersecurity topics within other computer science courses can increase students' awareness and knowledge regarding cybersecurity concepts. In contrast, we did not find significant knowledge improvements over time, as shown in the results of consecutive testing courses. We encourage higher education institutes to integrate some important cybersecurity topics within existing courses based on the obtained results. This adoption will significantly reduce security threats in different aspects, increase students' awareness of various cybersecurity topics, and enrich their knowledge.

### ACKNOWLEDGEMENT

The authors are grateful to the Applied Science Private University, Amman-Jordan, for the full financial support granted to cover the publication fee of this research article.